\documentclass[article,preprint,amssymb,showpacs,superscriptaddress]{revtex4}

\usepackage{graphicx}
\usepackage{dcolumn}
\usepackage{bm}
\usepackage{color}
\begin{document}

\renewcommand{\baselinestretch}{2} 

\title{Evidence for two distinct anisotropies in the oxypnictide superconductors SmFeAsO$_{0.8}$F$_{0.2}$ and NdFeAsO$_{0.8}$F$_{0.2}$}

\author{S.~Weyeneth} \email{wstephen@physik.uzh.ch}
\affiliation{Physik-Institut der Universit\"{a}t Z\"{u}rich, Winterthurerstrasse 190, CH-8057 Z\"urich, Switzerland}

\author{R.~Puzniak}
\affiliation{Institute of Physics, Polish Academy of Sciences, Aleja Lotnik\'ow 32/46, PL 02-668 Warsaw, Poland}

\author{N.D.~Zhigadlo}
\affiliation{Laboratory for Solid State Physics, ETH Zurich, CH-8093 Zurich, Switzerland}

\author{S.~Katrych}
\affiliation{Laboratory for Solid State Physics, ETH Zurich, CH-8093 Zurich, Switzerland}

\author{Z.~Bukowski}
\affiliation{Laboratory for Solid State Physics, ETH Zurich, CH-8093 Zurich, Switzerland}

\author{J.~Karpinski}
\affiliation{Laboratory for Solid State Physics, ETH Zurich, CH-8093 Zurich, Switzerland}

\author{H.~Keller}
\affiliation{Physik-Institut der Universit\"{a}t Z\"{u}rich, Winterthurerstrasse 190, CH-8057 Z\"urich, Switzerland}

\newpage
\begin{abstract}
Single crystals of the oxypnictide superconductors SmFeAsO$_{0.8}$F$_{0.2}$ and NdFeAsO$_{0.8}$F$_{0.2}$ with $T_\mathrm{c}$ in the range of 44 K to 48 K were investigated by torque magnetometry. An analysis of the data in terms of a recently proposed model for the anisotropic magnetization in the superconducting state, treating the magnetic penetration depth anisotropy $\gamma_\lambda$ differently than the upper critical field anisotropy $\gamma_H$, provides evidence that in the oxypnictide superconductors two distinct anisotropies are present. As a result $\gamma_\lambda$ differs significantly in magnitude and in temperature dependence from $\gamma_H$, analogous to MgB$_2$ but with a reversed sign of slope. This scenario strongly suggests a new multi-band mechanism in the novel class of oxypnictide high-temperature superconductors. 
\end{abstract}

\maketitle
It is well known, that all superconductors with the highest transition temperatures have a layered crystal structure, which manifests itself in pronounced anisotropic properties. In the recently discovered oxypnictide superconductors RFeAsO$_{1-x}$F$_x$ (R = La, Sm, Ce, Nd, Pr, Gd) with considerably high transition temperatures up to $T_\mathrm{c}\simeq56$ K superconductivity takes place in the FeAs layers, and the LaO sheets are charge reservoirs when doped with F ions \cite{Kamihara, Chen1, Chen2, Ren1, Ren2, Ren3, Cheng}. Whereas the cuprates have been characterized by a well-defined effective mass anisotropy, the observation of two distinct anisotropies in MgB$_2$ challenged the understanding of anisotropic superconductors \cite{Angst, Angstgap, Gurevich}. Therefore, it is  important to investigate the anisotropic behavior of the oxypnictides in order to clarify the nature of superconductivity in this novel class of superconductors. In this respect a detailed knowledge of the anisotropy parameter $\gamma$ is essential.

In the framework of phenomenological Ginzburg-Landau theory the anisotropic behavior of superconductors is described by means of the effective mass anisotropy parameter \cite{Kogan1}
\begin{equation}
\gamma=\sqrt{m^*_{c}/m^*_{ab}}=\lambda_c/\lambda_{ab}=\xi_{ab}/\xi_{c}=H_\mathrm{c2}^{||ab}/H_\mathrm{c2}^{||c}.
\label{Gamma}
\end{equation}
Here $m^*_{ab}$ and $m^*_{c}$ are the components of the effective carrier mass related to supercurrents flowing in the $ab$-planes and along the $c-$axis, respectively, $\lambda_{ab}$, $\lambda_c$, $\xi_{ab}$, and $\xi_c$ are the corresponding magnetic penetration depth and coherence length components, and $H_\mathrm{c2}^{||ab}$ and $H_\mathrm{c2}^{||c}$ the upper critical field components. For the oxypnictides many different estimates of the effective mass anisotropy with values ranging from 1 to 30 were reported \cite{Bernhard, Weyeneth, Prozorov, Jia2, Welp, Balicas, Kubota,Jaroszynski}. The first temperature dependent study of the anisotropy parameter $\gamma$ was performed on SmFeAsO$_{0.8}$F$_{0.2}$ single crystals by means of torque magnetometry \cite{Weyeneth}, where a strongly temperature dependent $\gamma$ was found, ranging from $\gamma\simeq8$ at $T\lesssim T_\mathrm{c}$ to $\gamma\simeq23$ at $T\simeq 0.4T_\mathrm{c}$. Other torque studies on SmFeAsO$_{0.8}$F$_{0.2}$ revealed a similar temperature dependence with $\gamma\simeq9$ saturating at lower temperatures \cite{Balicas}, whereas surprisingly for PrFeAsO$_x$ an almost temperature independent $\gamma\simeq1.1$ was reported \cite{Kubota}. A relatively low value of $\gamma\simeq2.5$ with a slight temperature dependence was also observed in magneto-optical studies performed on PrFeAsO$_x$ \cite{Koncykowski}. From upper critical field measurements, various investigations on oxypnictides have shown, that sufficiently below $T_\mathrm{c}$, $\gamma$ decreases with decreasing temperature, in sharp contrast to the magnetic torque data \cite{Weyeneth, Balicas} which reveal an increasing $\gamma$. A recent high magnetic field investigation \cite{Jaroszynski} of the upper critical field in single crystal NdFeAsO$_{0.7}$F$_{0.3}$ with a $T_\mathrm{c}=46$ K provided an estimate of the anisotropy $H_{c2}^{||ab}/H_{c2}^{||c}$ with $\gamma$ ranging from $\gamma\simeq6$ at $T\simeq38$ K to $\gamma\simeq5$ at $T\simeq34$ K. In summary, all these different estimates for $\gamma$ seem to be very difficult to understand in a consistent way in the framework of classical Ginzburg-Landau theory.

In order to clarify this puzzling behavior of the anisotropy parameter $\gamma$ we decided to perform further detailed torque studies on single crystals of nominal composition SmFeAsO$_{0.8}$F$_{0.2}$ and NdFeAsO$_{0.8}$F$_{0.2}$. The magnetic torque $\vec{\tau}$ of a sample with magnetic moment $\vec{m}$ in a magnetic field $\vec{H}$ is defined by
\begin{equation}
\vec{\tau}=\mu_0(\vec{m}\times \vec{H}).
\label{Deftorque}
\end{equation}
For anisotropic superconductors in the mixed state the diamagnetic magnetization is not strictly anti-parallel to the applied magnetic field due to the presence of vortices which are tilted in an arbitrary applied magnetic field. Therefore, an anisotropic superconductor in a magnetic field will exhibit a magnetic torque according to Eq.~(\ref{Deftorque}). In the mean-field approach of the anisotropic Ginzburg-Landau theory the torque for a superconductor with a single gap is written as \cite{Kogan1}
\begin{equation}
\tau(\theta)=-\frac{V\Phi_0H}{16\pi\lambda_{ab}^2}\Bigg(1-\frac{1}{\gamma^2}\Bigg)\frac{\sin(2\theta)}{\epsilon(\theta)}\ln\Bigg(\frac{\eta H_\mathrm{c2}^{||c}}{\epsilon(\theta)H}\Bigg).
\label{kogan}
\end{equation}
$V$ is the volume of the crystal, $\Phi_0$ is the elementary flux quantum, $H_{c2}^{||c}$ is the upper critical field along the $c$-axis of the crystal, $\eta$ denotes a numerical parameter of the order unity, depending on the structure of the flux-line lattice, and $\epsilon(\theta)=[\cos^2(\theta)+\gamma^{-2}\sin^2(\theta)]^{1/2}$. Three fundamental thermodynamic parameters can be extracted from the angular dependence of the torque in the mixed state of a superconductor: the in-plane magnetic penetration depth $\lambda_{ab}$, the $c$-axis upper critical field $H_{c2}^{||c}$, and the effective mass anisotropy $\gamma$. As has been pointed out by Kogan \cite{Kogan2}, the anisotropies for the magnetic penetration depth $\gamma_\lambda=\lambda_{c}/\lambda_{ab}$ and for the upper critical field $\gamma_H=H_\mathrm{c2}^{||ab}/H_\mathrm{c2}^{||c}$, do not necessarily coincide for unconventional Ginzburg-Landau superconductors as described in Eq.~(\ref{Gamma}) where $\gamma=\gamma_\lambda=\gamma_H$. A more generalized approach including the two distinct anisotropies $\gamma_\lambda$ and $\gamma_H$, leads to the more general expression \cite{Kogan2}
\begin{equation}
\tau(\theta)=-\frac{V\Phi_0H}{16\pi\lambda_{ab}^2}\Bigg(1-\frac{1}{\gamma_\lambda^2}\Bigg)\frac{\sin(2\theta)}{\epsilon_\lambda(\theta)}\Bigg[\ln\Bigg(\frac{\eta H_\mathrm{c2}^{||c}}{H}\frac{4e^2\epsilon_\lambda(\theta)}{(\epsilon_\lambda(\theta)+\epsilon_H(\theta))^2}\Bigg)-\frac{2\epsilon_\lambda(\theta)}{\epsilon_\lambda(\theta)+\epsilon_H(\theta)}\Bigg(1+\frac{\epsilon'_\lambda(\theta)}{\epsilon'_H(\theta)}\Bigg)\Bigg].
\label{kogan2}
\end{equation}
Here the scaling function $\epsilon_i(\theta)=[\cos^2(\theta)+\gamma_i^{-2}\sin^2(\theta)]^{1/2}$ with $i=\lambda,H$ is different for $\gamma_\lambda$ and $\gamma_H$. $\epsilon'_i(\theta)$ denotes its derivative with respect to the angle $\theta$.

\begin{figure}[t!]
\includegraphics[width=1\linewidth]{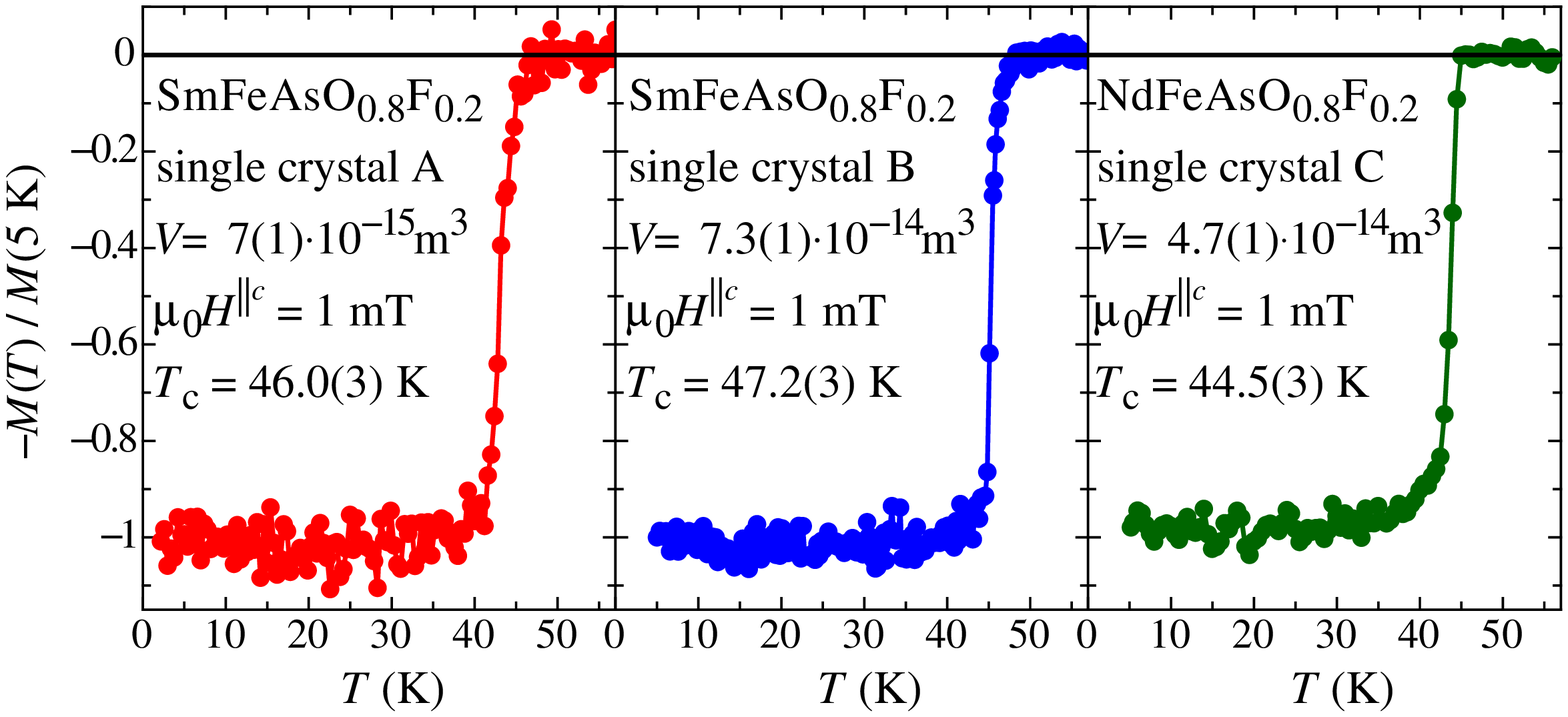}
\vspace{0cm}
\caption{(color online) Normalized magnetization for the three single crystals studied in this work. The magnetic moment was measured in the zero field cooling mode with an applied field of 1 mT parallel to the $c$-axis. Below $T_\mathrm{c}$ all samples show full diamagnetic response with a narrow and well defined transition temperature.}
\label{fig1}
\end{figure}

The method of crystal growth and the basic superconducting properties of the SmFeAsO$_{1-x}$F$_y$ single crystals investigated here were already reported \cite{Karpinski2}. The same procedure was used to grow single crystals of NdFeAsO$_{1-x}$F$_y$. The plate-like crystals used in this work were of rectangular shape with typical masses of the order of 100 ng. The crystal structure was checked by means of X-ray diffraction revealing the $c$-axis to be perpendicular to the plates. The magnetization curves shown in Fig.~\ref{fig1} were measured in the Meissner state using a commercial Quantum Design SQUID magnetometer MPMS XL with installed Reciprocating Sample Option. Small variation of the transition temperature of the various samples may be due to oxygen and fluorine deficiency. The volume of the crystals given in Fig.~\ref{fig1} was estimated from magnetization measurements in low magnetic fields applied along the samples $ab$-plane. The $c$-axis dimension is much smaller than the $a$- and $b$-dimensions, and therefore demagnetizing effects can be neglected. Good agreement was found with optical microscope measurements of the dimensions of the sample. 

\begin{figure}[t!]
\includegraphics[width=1\linewidth]{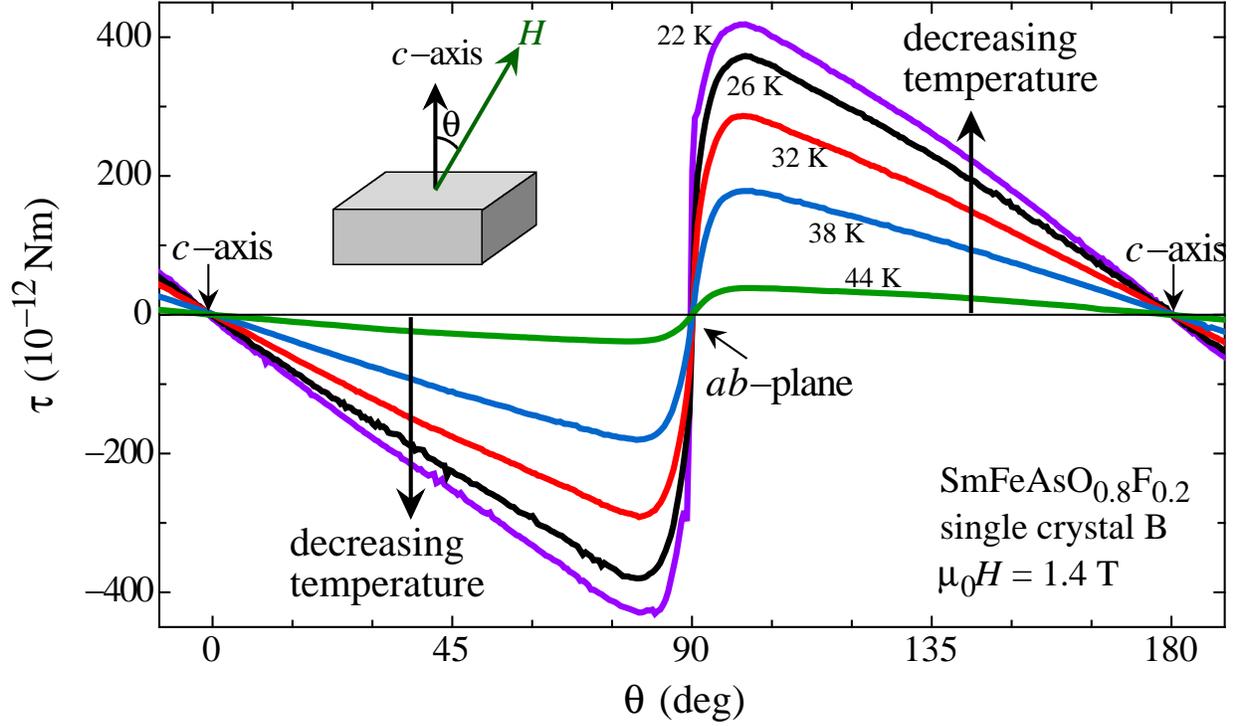}
\vspace{0cm}
\caption{(color online) Angular dependence of the reversible torque data for SmFeAsO$_{0.8}$F$_{0.2}$ (single crystal B) at several temperatures derived in a magnetic field of 1.4 T. Only a small almost temperature independent background contribution smaller than 10$^{-11}$ Nm is present in the data, stemming from a minor anisotropic normal state magnetization. For clarity not all measured data are shown.}
\label{fig2}
\end{figure}

To perform most accurate measurements of the superconducting anisotropy parameter, we have chosen high sensitivity torque magnetometry. For the low field torque measurements we used an experimental set-up described in detail elsewhere \cite{Weyeneth}. It is worth pointing out that the anisotropy data published several years ago for single-crystal MgB$_2$ were also obtained with this torque set-up \cite{Angst}. In order to derive the reversible torque  from the raw data, recorded in the irreversible regime by clockwise ($\theta^+$) and counter-clockwise ($\theta^-$) rotating of the magnetic field with respect to the $c$-axis, we used the standard approximation $\tau(\theta)=(\tau_{raw}(\theta^+)+\tau_{raw}(\theta^-))/2$ described elsewhere \cite{Weyeneth}. Some of the reversible torque data for SmFeAsO$_{0.8}$F$_{0.2}$ (single crystal B) are shown in Fig.~\ref{fig2}.

\begin{figure}[t!]
\includegraphics[width=1\linewidth]{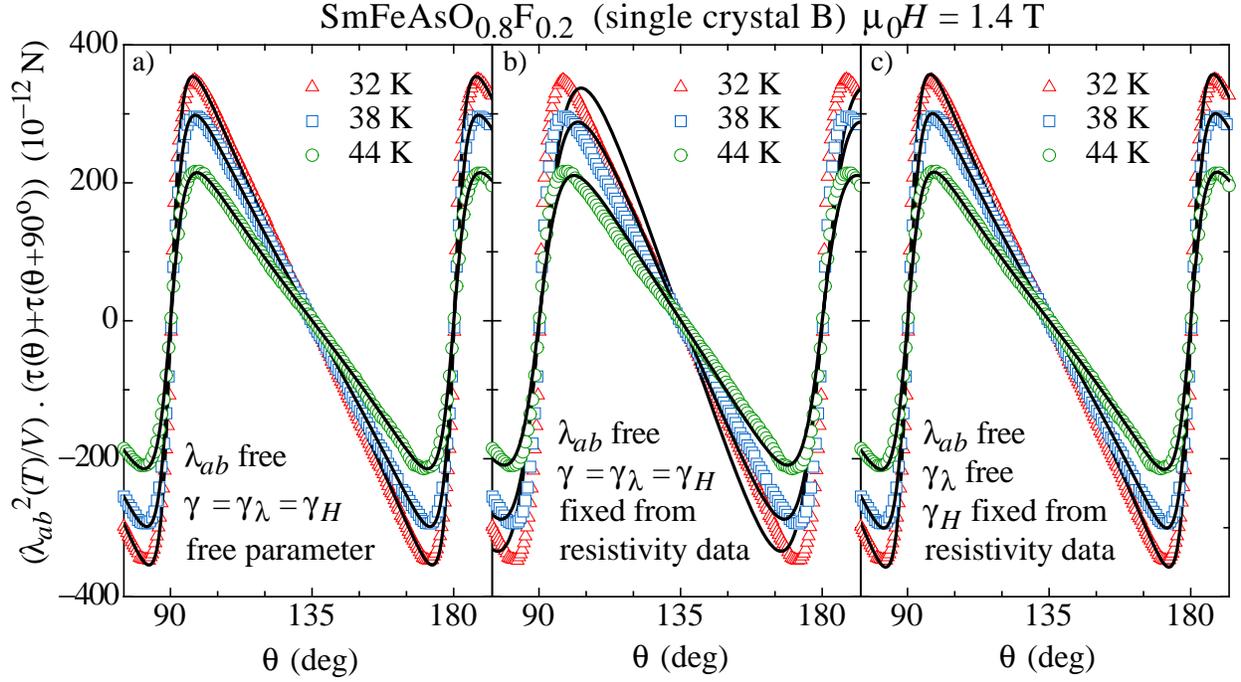}
\vspace{0cm}
\caption{(color online) Three sets of normalized torque data, obtained at 32 K, 38 K, and 44 K in 1.4 T for SmFeAsO$_{0.8}$F$_{0.2}$ (single crystal B) and analyzed in terms of $\tau(\theta)+\tau(\theta+90^\circ)$. a) Data described with Eq.~(\ref{kogan}) with both $\gamma$ and $\lambda_{ab}$ as free parameters. b) Data described with Eq.~(\ref{kogan}) with $\gamma$ fixed to the linearly extrapolated values of the resistivity measurements \cite{Jaroszynski} and with $\lambda_{ab}$ as a free parameter. c) Data described with the generalized Eq.~(\ref{kogan2}) with $\gamma_H$ fixed to the linearly extrapolated values of the resistivity measurements \cite{Jaroszynski} and with $\gamma_\lambda$ and $\lambda_{ab}$ as free parameters.}
\label{fig3}
\end{figure}

\begin{figure}[t!]
\includegraphics[width=0.8\linewidth]{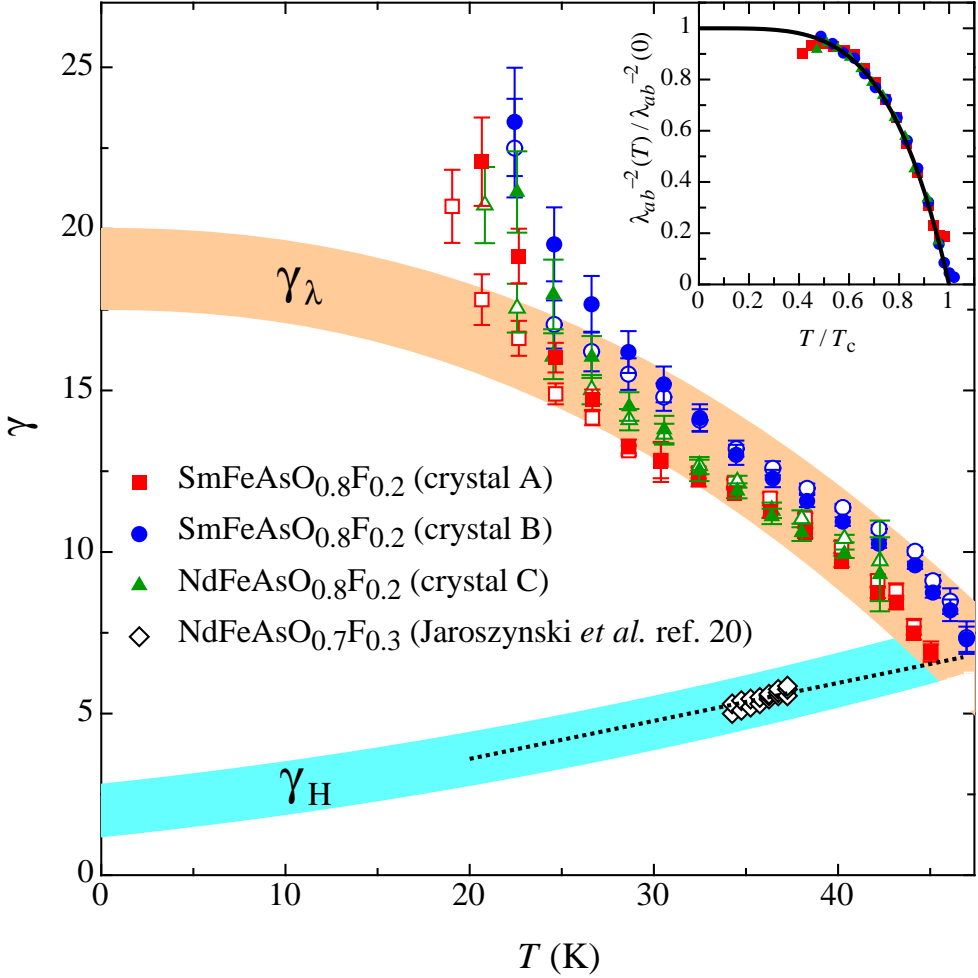}
\vspace{0cm}
\caption{(color online) Summary of all the parameters derived from the systematic analysis of the torque data of single crystals SmFeAsO$_{0.8}$F$_{0.2}$ (crystal A and B) and NdFeAsO$_{0.8}$F$_{0.2}$ (crystal C). Values of $\gamma$ (\textit{open symbols}) and $\gamma_\lambda$ (\textit{closed symbols}) were obtained from fits to the data with Eqs.~(\ref{kogan}) and (\ref{kogan2}), respectively. The upper \textit{broad orange band} is a guide to the eye, suggesting an estimate of $\gamma_\lambda(0)\approx19$. The \textit{dotted line} is the linear extrapolation of $\gamma_H$ obtained from resistivity measurements \cite{Jaroszynski} on a NdFeAsO$_{0.7}$F$_{0.3}$ single crystal with similar $T_\mathrm{c}=45$ K (\textit{diamonds}). The lower \textit{broad blue band} is a guide to the eye, suggesting an estimate of $\gamma_H(0)\approx2$. The inset shows the normalized superfluid density as obtained from fits of Eq.~(\ref{kogan2}) to the torque data. The \textit{solid line} denotes the best fit to the data using the power law in Eq.~(\ref{superfluid}) as explained in the text.}
\label{fig4}
\end{figure}

\begin{figure}[t!]
\includegraphics[width=0.8\linewidth]{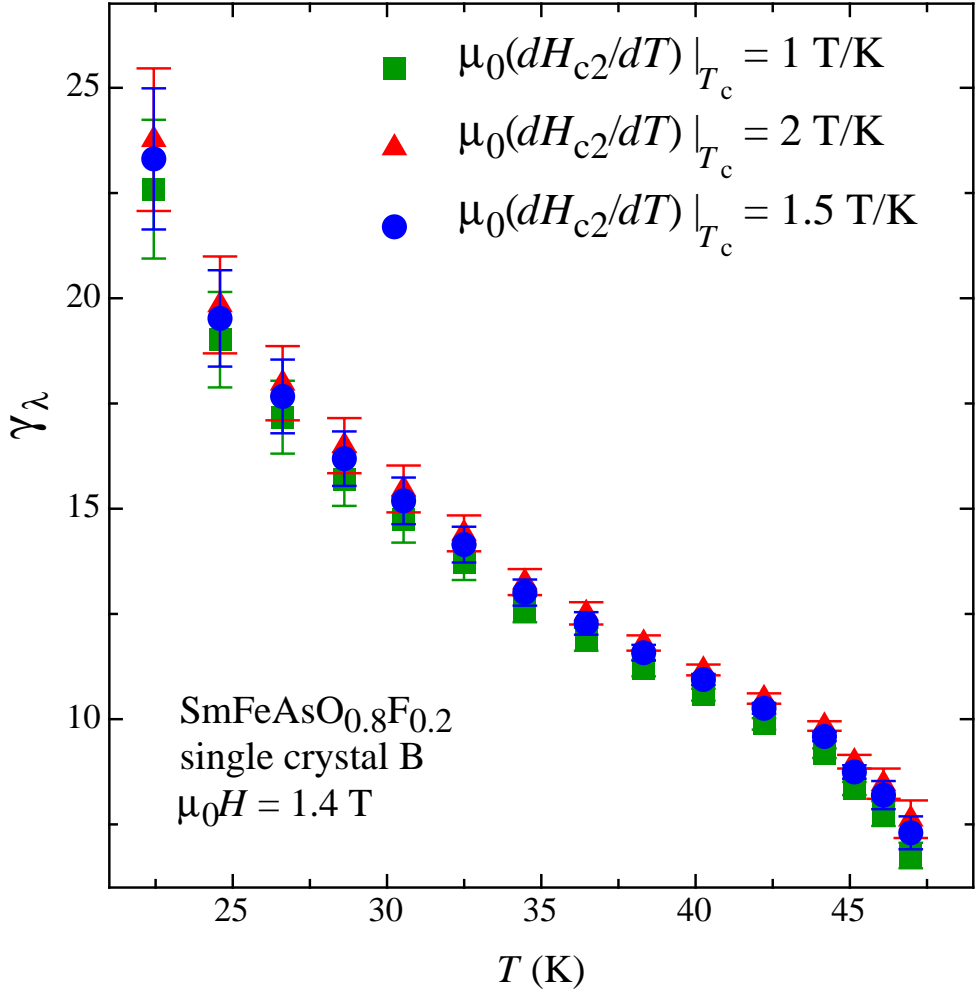}
\vspace{0cm}
\caption{(color online) Values of $\gamma_\lambda$ for SmFeAsO$_{0.8}$F$_{0.2}$ (crystal B) obtained from fits to the data with Eq.~(4). The used WHH relation for $H_\mathrm{c2}^{||c}(T)$ was varied assuming three different slopes $\mu_0(dH_\mathrm{c2}^{||c}/dT)|_{T_\mathrm{c}}=1$ T/K, 1.5 T/K and 2 T/K as explained in the text. The resulting temperature dependencies of $\gamma_\lambda$ are within experimental errors the same, and therefore do not depend significantly on the slope. The anisotropy can be reliably determined by fixing the slope to 1.5 T/K, which is a typical value reported for oxypnictide superconductors.}
\label{fig5}
\end{figure}

The torque data in the superconducting state were analyzed using the approach proposed by Balicas \emph{et al.} \cite{Balicas} in order to eliminate any anisotropic paramagnetic or diamagnetic background contribution. In this approach the torque data were first symmetrized according to $\tau_{symm}(\theta)=\tau(\theta)+\tau(\theta+90^\circ)$ and subsequently analyzed. In the fitting procedure we always kept the upper critical field fixed, assuming a WHH dependence \cite{WHH} with a slope at $T_\mathrm{c}$ of $\mu_0(dH_\mathrm{c2}^{||c}/dT)|_{T_\mathrm{c}}$=1.5 T/K, which is a typical value reported for oxypnictide superconductors \cite{Weyeneth,Hunte, Jaroszynski, Jia2, Welp}. This approach reduces the number of free fit parameters, so that both $\gamma$ and $\lambda_{ab}$ can be reliably determined in a simultaneous fit. By fitting the simple Kogan expression in Eq.~(\ref{kogan}) to the angular dependent torque data, we get excellent agreement as shown in Fig.~\ref{fig3}a. The resulting values of $\gamma$ determined for all samples are shown in Fig.~\ref{fig4}. It is important to stress, that the choice of the actual slope $\mu_0(dH_\mathrm{c2}^{||c}/dT)|_{T_\mathrm{c}}$=1.5 T/K  doesn't influence the determination of $\gamma$ much, since $\gamma$ is mostly sensitive to the angular dependence of the torque close to the $ab$-plane. We tested the influence of a variation of $H_\mathrm{c2}^{||c}(T)$ on the error in $\gamma$ by altering the slope $\mu_0(dH_\mathrm{c2}^{||c}/dT)|_{T_\mathrm{c}}$ from 1 T/K to 2 T/K. As a result, $\gamma$ varies only within a few percent as depicted in Fig. \ref{fig5}. All of the anisotropy data obtained with different slopes are the same within the experimental errors given in Fig.~\ref{fig4}. With $H_\mathrm{c2}^{||c}$ as a free parameter the scattering of all fitted parameters is strongly enhanced, and unphysical values for the upper critical field are obtained, which was already noted in earlier torque studies \cite{Weyeneth, Balicas, Kubota}.

The analysis of the torque data presented so far leads to a pronounced temperature dependent $\gamma$ with values of $\gamma\geqslant20$ for $T\lesssim20$ K \cite{Weyeneth}. However, this result is in strong contrast to the results obtained from resistivity data on single crystals of the similar compound NdFeAsO$_{0.7}$F$_{0.3}$ where the anisotropy $H_\mathrm{c2}^{||ab}/H_\mathrm{c2}^{||c}$  was found to decrease with decreasing temperature \cite{Jaroszynski}. Different techniques should lead to similar values for the same quantity, and therefore it is important to explain the physical meaning of this discrepancy. In Fig.~\ref{fig3}b we depict the same data set shown in Fig.~\ref{fig3}a, analyzed according to Eq.~(\ref{kogan}) with $\gamma$ fixed to linearly extrapolated values of the resistivity measurements of Jaroszynski \emph{et al.} \cite{Jaroszynski} and only with $\lambda_{ab}$ as a free parameter. Obviously, the fit does not describe the torque data well. This discrepancy suggests the presence of two distinct anisotropies, namely $\gamma_\lambda=\lambda_{c}/\lambda_{ab}$ and $\gamma_H=H_\mathrm{c2}^{||ab}/H_\mathrm{c2}^{||c}$. In order to test this hypothesis we analyzed the torque data with the generalized expression Eq.~(\ref{kogan2}), including both $\gamma_{\lambda}$ and $\gamma_H$. The result is shown in Fig.~\ref{fig3}c, again with the same fixed $\gamma_H$ as in Fig.~\ref{fig3}b, but with $\gamma_{\lambda}$ as a free parameter. Note that this approach describes all torque data very well. Moreover, the values of $\gamma_{\lambda}$ are very similar to those previously obtained by means of Eq.~(\ref{kogan}) \cite{Weyeneth}. The final results of fitting $\gamma_\lambda$ and $\lambda_{ab}$ by means of Eq.~(\ref{kogan2}) are shown in Fig.~\ref{fig4}. In the temperature range $T_\mathrm{c}\geqslant T\geqslant25$ K, all values for $\gamma_\lambda$ show the same temperature behavior, which appears to be generic for oxypnictide superconductors. From this trend a value of $\gamma_\lambda(0)\approx19$ can be estimated. The drastic increase of $\gamma_\lambda$ below 25 K to values well above 20, is very likely due to a substantial pinning contribution to the torque which makes it impossible to derive reversible torque data from measurements in the irreversible region. Moreover, in this regime the fit parameters deviate from each other and depend even on the used fitting expression, further supporting the presence of non-equilibrium effects, such as strong pinning. The pronounced temperature dependence of $\gamma_\lambda$ between 25 K and $T_\mathrm{c}$ strongly signals multi-band superconductivity, like in e.g. MgB$_2$. However, in order to clarify its origin more systematic experimental work is required.

Further support for the proposed scenario stems from the temperature dependence of the in-plane magnetic penetration depth $\lambda_{ab}$. In the inset to Fig.~\ref{fig4} the normalized superfluid density $\lambda_{ab}^{-2}(T)/\lambda_{ab}^{-2}(0)$ as obtained from fits to Eq.~(\ref{kogan2}) is shown for the three single crystals (A, B, C). It is well described by the power law
\begin{equation}
\lambda_{ab}^{-2}(T)/\lambda_{ab}^{-2}(0)=1-\big(T/T_c\big)^n
\label{superfluid}
\end{equation}
with $n=4.2(3)$, which is close to $n=4$ characteristic for a superconductor in the very strong-coupling limit \cite{Rammer}. The zero temperature value of $\lambda_{ab}(0)\approx250(50)$ nm obtained for all samples is in reasonable agreement with other values reported \cite{Luetkens, Khasanov, Weyeneth, Malone}.

It is evident from Fig.~\ref{fig4} that two distinct anisotropies $\gamma_\lambda$ and $\gamma_H$ describe the torque data consistently. Equation~(\ref{kogan2}) was originally proposed for the two-band superconductor MgB$_2$. Indeed the experimental situation in both materials is quite similar. MgB$_2$ shows two distinct anisotropies: $\gamma_\lambda$ decreases with decreasing temperature from about 2 to 1.1, whereas  $\gamma_H$ increases from about 2 at $T_c$ to values up to 6 at low temperatures \cite{Angst, Angstgap, Gurevich}. For the oxypnictide superconductors investigated here a similar situation appears to be present, but with reversed signs of the slopes of $\gamma_\lambda(T)$ and $\gamma_H(T)$. In MgB$_2$ the existence of two distinct bands of different dimensionality, and with strong interband and intraband scattering was suggested to be responsible for the high $T_\mathrm{c}$'s \cite{Angstgap, Bussmann}.

In conclusion, we found strong evidence that two distinct anisotropies $\gamma_\lambda$ and $\gamma_H$ are involved in the superconductivity of the oxypnictide superconductors. Torque magnetometry in low magnetic fields is very sensitive to $\gamma_\lambda$ which exhibits a pronounced increase with decreasing temperature. This is in contrast to the behavior of $\gamma_H$ which according to recent resistivity measurements decreases with decreasing temperature \cite{Jaroszynski}. Close to $T_\mathrm{c}$ both anisotropies have very similar values of $\gamma_\lambda(T_\mathrm{c})\approx\gamma_H(T_\mathrm{c})\approx7$, whereas at low temperatures the magnetic penetration depth anisotropy $\gamma_\lambda(0)\approx19$ and the upper critical field anisotropy $\gamma_H(0)\approx2$. This behavior is similar to the situation in the two-band superconductor MgB$_2$ where the temperature dependencies of the two anisotropies are well understood \cite{Angstgap,Gurevich,Bussmann}. This result strongly suggests multi-band superconductivity in the novel class of oxypnictide superconductors, as already suggested in previous investigations \cite{Hunte, Weyeneth, Balicas, Malone}.

\section{Acknowledgments}
The authors are grateful to B. Graneli for the help to prepare the manuscript. This work was supported by the Swiss National Science Foundation and in part by the NCCR program MaNEP, the Polish Ministry of Science and Higher Education within the research project for the years 2007 - 2009 (No.~N~N202~4132~33), and the EU Project CoMePhS.

\end{document}